\documentclass[12pt]{article}
\usepackage{epsf}
\usepackage[dvips]{graphicx}

\let\ensm=\ensuremath
\def\Pp{{\rm p}}

\def\PK{{\rm K}}

\def\Pu{{\rm u}}
\def\Pd{{\rm d}}
\def\Ps{{\rm s}}

\def\pip               {\ensm{\pi^+}}
\def\pim               {\ensm{\pi^-}}
\def\pipm              {\ensm{\pi^\pm}}

\def\gev              {\ensm{\,{\rm GeV}}}
\def\gevp             {\ensm{\,{\rm GeV}/c}}

\begin{document}

\title{Asymmetry measurement of charged hadron production in
 p$\uparrow$A collisions at 40 GeV}

\author{V.~Abramov, P.~Goncharov, A.~Kalinin, A.~Khmelnikov,\\
 A.~Korablev, Yu.~Korneev, A.~Kostritski,  A.~Markov, \\
A.~Krinitsyn, V.~Kryshkin, V.~Talov, L.~Turchanovich, A.~Volkov }

\maketitle
\vskip 1mm

\begin{center}

{\small
(1) {\it
Institute for High Energy Physics, Protvino, Moscow region, Russia
}
\\
$\dag$ {\it
E-mail: Victor.Kryshkin@ihep.ru
}}
\end{center}


\vskip 10mm

\centerline{\bf Abstract}

{The single spin asymmetry of charge hadron production by a 40 $\gev$
 proton beam with 39\% transverse polarization incident on nuclei (C, Cu)
 has been measured
using the Focusing Double Arm Spectrometer (FODS). The measurements were
carried out for hadrons with high $x_T$ ($82^{\circ}$  in c.m.) and with
high $x_F$ (one arm at $51^{\circ}$  and the other arm 
at $99^{\circ}$  in c.m.). 
 The results are presented for charged pions, kaons, protons and
antiprotons with high $x_T$. }


%
\vskip 30mm
\begin{center}
{\it   Talk delivered
 by V. Kryshkin at the ``16th International Spin Physics Symposium'', SPIN2004,
 October 10-16, 2004, Trieste, Italy}
\end{center}
\newpage

We have measured the single-spin asymmetry $A_N$ of the inclusive charged
pion, kaon, proton and antiproton production cross sections at high
$x_T$ and high $x_F$ for a 40 $\gevp$ proton beam incident on nuclei (C, Cu), 
where $A_N$ is defined as
 \begin{equation}
 A_{N} = { 1\over{P_{B}\cdot cos{\phi}} } \cdot
{ {N{\uparrow} - N{\downarrow}} \over {N{\uparrow} + N{\downarrow}} }, \quad
\end{equation}
where $P_B$ is the beam polarization, $\phi$ is the athimuthal angle of the 
production plane, ${N\uparrow}$ and ${N\downarrow}$ are event rates
for the beam spin up and down respectively. The measurements were
carried out at IHEP, Protvino. The 
polarized protons are produced by the parity - nonconserving $\ensm{\Lambda}$
decays [1].
 The up or down beam transverse
polarization is achieved by the selection of decay protons with angles
near $90^{\circ}$ in the $\ensm{\Lambda}$ rest frame by
a movable collimator.  At the end of the beam line two  magnets 
 correct the vertical beam position on the spectrometer target for the two
beam polarizations. The intensity of the 40 $\gevp$ momentum
polarized beam on the spectrometer target  is $3\times 10^{7}$ ppp,
 $\Delta p/p= \pm 4.5$\%, the transverse polarization is $39^{+1}_{-3}$ \%,
 and the polarization direction
is changed each 18 min during 30 s. The beam intensity and position
are measured by  ionization chambers and  scintillation
hodoscopes. Two Cherenkov counters 
 identify the beam particle
composition to control background contamination.  At the spectrometer
magnet entrance there are two scintillation hodoscopes to measure
the vertical coordinates of particles emitted from the target.

  The measurements have been carried out with the FODS [1] spectrometer.
  It consists of an analyzing magnet,  drift chambers, the
  Cherenkov radiation spectrometer (SCOCH) for particle
identification ($\pipm$ , $\PK^{\pm}$ , $\Pp$ and $\bar{\rm{p}}$), 
 scintillation counters and 
hadron calorimeters to  trigger on the high energy hadrons. To
further suppress a background there are two threshold Cherenkov
counters using air at atmospheric pressure inserted in the
magnet. Inside the magnet there is also a beam dump made of tungsten
and copper.  There are two arms which can be rotated around the target
center situated in front of the magnet to change the secondary
particle angle.
  The  Cherenkov radiation spectrometer
consists of a spherical mirror with diameter 110 cm, 24
cylindrical lenses to focus the Cherenkov light on the hodoscope
photomultipliers. Measuring the particle velocity using the SCOCH and
its momentum in the magnetic field one can determine the particle square
mass $\ensm{\,{\rm M}}^2$.  The SCOCHs are filled with
Freon~13 at 8~atm.  

In 1994 a study of the single spin asymmetry
($A_N$) in  inclusive charge hadron production was started using FODS:
\begin{equation}
\rm{p\uparrow + p(A) \rightarrow h^{\pm} + X},
\end{equation}
\begin{equation}
\rm{p\uparrow + p(A) \rightarrow h^{\pm}+ h^{\pm}  + X},
\end{equation}
 where $\rm{h^{\pm}}$ is a charged hadron
(pion, kaon, proton or antiproton).  The experimental program consists of 
 measuring the
charge hadron single spin asymmetry at high $x_T$ and $x_F$ in
$\Pp\Pp$ and $\Pp$A collisions to study the
asymmetry dependence on the quark flavors $\Pu$, $\Pd$, $\Ps$
 and kinematical variables.

The pilot measurements of $A_N$ for the charged hadrons carried out in
1994 on a hydrogen target for small $x_F$ [1]
 are presented below for
comparison with data obtained with  nuclei. 

 The measurements of $A_{N}$ in
the range $-0.15 \le x_{F} \le 0.2$ and $0.5 \le p_{T} \le 4$ GeV/$c$ 
are carried out with symmetrical
arm positions at  angles of $\pm$ 160 mrad. The results of the two arms are
averaged, which partially cancels systematical uncertaities connected with the
variation of the beam position in the vertical direction, the
intensity monitor and the apparatus  drift.
\begin{figure}[ht]
\centerline{\epsfxsize=6.0in\epsfbox{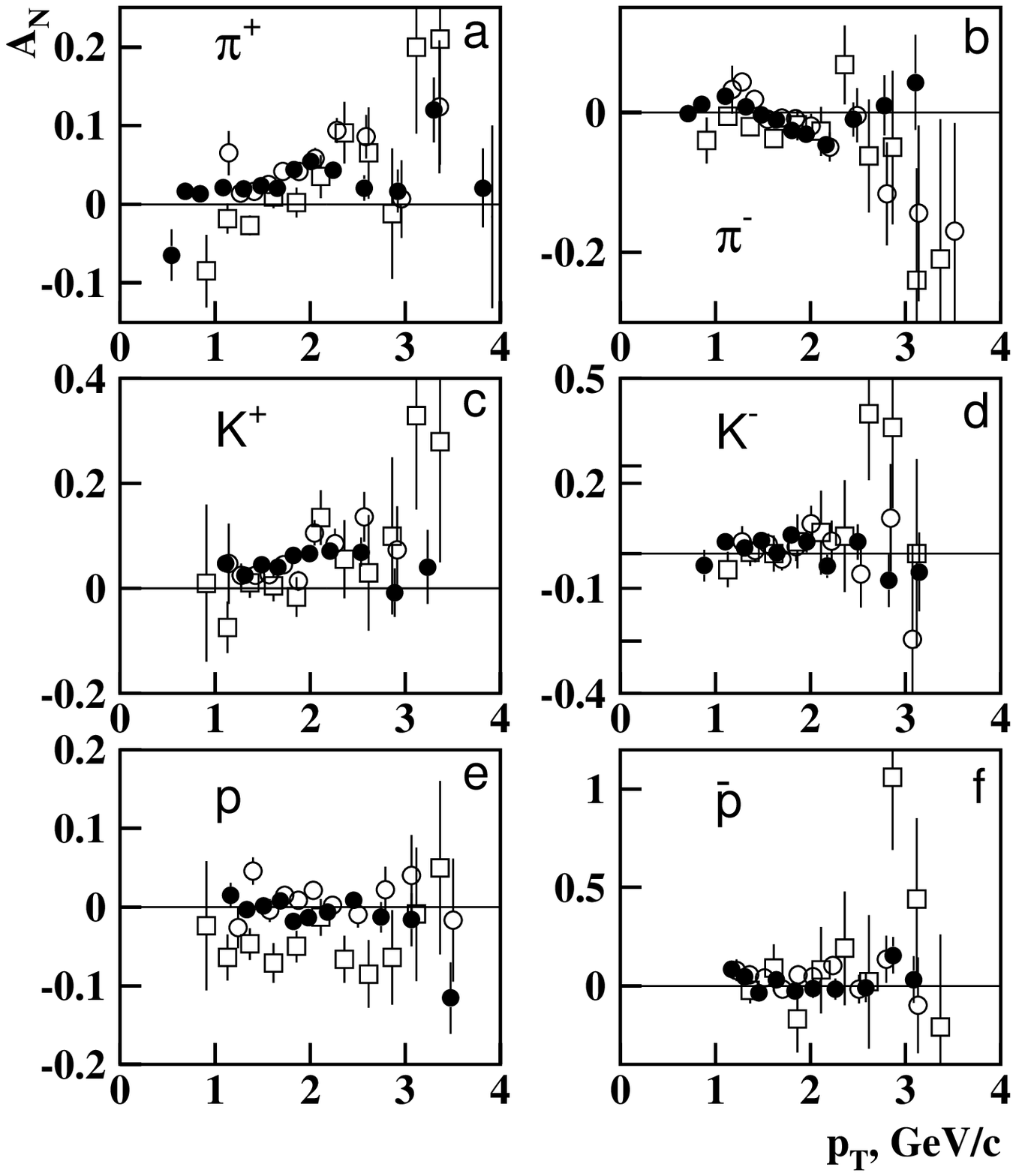}}   
\caption{$A_{N}$ dependence on  $p_T$ for 
$\rm{p\uparrow + p(A) \rightarrow h^{\pm} + X}$, where $\rm{h=\pip}$ (a),
 $\rm{\pim}$ (b),  $\rm{\PK^{+}}$ (c), $\rm{\PK^{-}}$ (d), $\Pp$ (e),
   $\rm{\bar{p}}$ (f). Closed cirles correspond to C target, open cirles - Cu,
 square - proton.   \label{asym}}
\end{figure}

  Figure~\ref{asym}a  depicts the $\pip$ 
meson production asymmetry.  
Within the errors there is no difference
of $A_{N}$ for  both targets (C and Cu). $A_{N}$  for the nuclear targets in
the range $1 \le p_{T} \le 2$ $\gevp$ is approximately 4\% higher than
 for the hydrogen
target. For the central region such a difference can be connected with
the smaller portion of $\Pu$ quarks in the nuclear target containing
neutrons. 
Fragmentation of $\Pu$ quarks 
($\rm{u}\rightarrow \pip $)
from the polarized beam protons as well as  $\Pu$ quarks of the target
contribute to the asymmetry.
 Because the target protons are not polarized their
contribution in the central region reduces the measured polarization.
For nuclear targets containing less $\Pu$ quarks in comparison with $\Pd$
quarks the decrease of the asymmetry is not so substantial. Quark
scattering in nuclei must also lead to the decrease of the asymmetry. 

 The asymmetry for $\pim$ meson production is presented 
in Figure~\ref{asym}b. In the range $0.9 \le p_{T}
\le 1.6$ $\gevp$ it is about 4\% higher for the nuclear targets than for the
hydrogen target. For the central region such differences can be
connected with the larger proportion of $\Pd$ quarks in the nuclear targets.
 The major  fragmentation contribution 
give $\Pd$ quarks ($\Pd \rightarrow \pim $) from polarized beam
 protons and the target. For $\pim$
mesons in $\Pp\Pp$ collisions the asymmetry is therefore negative. 
Due to the large
contribution of the unpolarized target in the central regions the asymmetry
for nuclear targets is shifted into the positive region. 

 Figure~\ref{asym}c shows
the asymmetry for $\PK^+$ production. There is no significant
difference in $A_N$ for the two
nuclear targets (C and Cu)  and $A_N$ is about
 3\% higher than for the hydrogen target. The reason for this can be 
the same as for  $\pip$ mesons.

Figure~\ref{asym}d presents $A_N$ for $\PK^-$ mesons.
 Within the errors there is no appreciable   difference in $A_{N}$ for
 all targets ($\Pp$, C and Cu) and $A_N$ is
 close to zero. This is expected because $\PK^-$ does not contain valence
 quarks from the beam proton.

Figure~\ref{asym}e depicts the asymmetry for  proton production
 which is  close to zero in nuclear targets.
For the hydrogen target it is slightly negative.

 The
asymmetry for antiproton production presented in Figure~\ref{asym}f shows no
difference for all targets ($\Pp$, C and Cu) and is close to zero.
This result is expected  because the produced antiproton
does not contain valence quarks from the beam proton. Sea quarks in
most models are expected to be unpolarized. The upper limit for the
asymmetry for antiprotons and $\PK^-$ mesons is given by the sensitivity
 of the experiment ( 4\%).

  Two features of the results can be stressed:

 1. There is no significant  difference for the two nuclear targets (C, Cu);

2. For the  positive charge mesons the asymmetry has a maximum at  $p_{T}=2.2$
$\gevp$ and decreases to zero at $p_{T}=2.9$ $\gevp$.

 The analysis for high $x_F$ is still under way.

\end{document}